\documentclass[12pt]{article}%
\usepackage{amsmath,latexsym}
\usepackage{graphicx}
\usepackage{amsmath}
\usepackage{amsfonts}
\usepackage{amssymb}%
\begin{document}

\title{{Some examples of Hayward wormholes}}
   \author{
  Peter K.F. Kuhfittig\\  \footnote{kuhfitti@msoe.edu}
 \small Department of Mathematics, Milwaukee School of
Engineering,\\
\small Milwaukee, Wisconsin 53202-3109, USA}

\date{}
 \maketitle

\begin{abstract}\noindent
The first part of this paper discusses a model for the
theoretical construction of a simple traversable
wormhole with zero density that depends on a preexisting
black hole.  By assuming the interconvertibility of
black holes and wormholes proposed by S.A. Hayward,
it is shown that a toy model suggested by the first
model may yield several possible transitions from the
preexisting black hole to a wormhole.  A final topic
is the conversion to a wormhole by assuming a specific
model for the exotic matter.\\

\noindent
PAC numbers: 04.20.Jb, 04.20.Gz
\end{abstract}

\section{Introduction}\label{E:introduction}

Wormholes are handles or tunnels in the spacetime
topology linking widely separated regions of our
Universe or of different universes altogether.  While
just as good a prediction of Einstein's theory as
black holes, they have so far eluded detection.
Moreover, holding a wormhole open requires a
violation if the null energy condition \cite{MT88}.

The first part of this paper discusses wormholes
that have zero density and must therefore depend
on preexisting black holes.  The interconvertibility
of black holes and wormholes proposed by Hayward
\cite{sH02, sH98} then leads to a toy model
suggested by the first model and which results in
several possible transitions from the preexisting
black hole to a wormhole.  The last model discussed
assumes a specific equation of state for the exotic
matter and shows that the singularity of the black
hole dissolves to become part of the wormhole's
mass.

To study the effects of a constant density, we
begin with the line element describing a static
spherically symmetric wormhole is given by
\cite{MT88} \begin{equation}\label{E:line1}
ds^{2}=-e^{2\Phi(r)}dt^{2}+\frac{dr^2}{1-b(r)/r}
dr^{2}+r^{2}(d\theta^{2}+\text{sin}^{2}\theta\,
d\phi^{2}).
\end{equation}
(We are using units in which $c=G=1$.)  Here $\Phi=
\Phi(r)$ is the \emph{redshift function}, which must
be everywhere finite to avoid an event horizon,
while $b=b(r)$ is the \emph{shape function}.  For the
shape function, $b(r_0)=r_0$, where $r=r_0$ is the
\emph{throat} of the wormhole.  A key requirement is
the \emph{flare-out condition}, $b'(r_0)<1$, which
indicates a violation of the null energy condition.
A concrete manifestation of this violation is the
need  for exotic matter in the vicinity of the throat.
In this region we also require that $b(r)<r$.

The Einstein field equations are stated next.
\begin{equation}\label{E:Einstein1}
   8\pi\rho=\frac{b'}{r^2},
\end{equation}
\begin{equation}\label{E:Einstein2}
   8\pi p_r=-\frac{b}{r^3}+2\left(1-\frac{b}{r}\right)
   \frac{\Phi'}{r},
\end{equation}
\begin{equation}\label{E:Einstein3}
   8\pi p_t=\left(1-\frac{b}{r}\right)\left(\Phi''-
   \frac{b'r-b}{2r(r-b)}\Phi'+(\Phi')^2+
   \frac{\Phi'}{r}-\frac{b'r-b}{2r^2(r-b)}\right).
\end{equation}
Comparing Eqs. (\ref{E:line1}) and (\ref{E:Einstein1}),
if $b(r)\equiv\text{constant}$, then $\rho(r)\equiv 0$.
That zero-density wormholes may be of interest is not
new.  Thus Visser \cite{mV96} discusses the wormhole
metric
\begin{equation}\label{E:line2}
ds^{2}=-e^{2\Phi(r)}dt^{2}+\frac{dr^2}{1-r_0/r}
+r^{2}(d\theta^{2}+\text{sin}^{2}\theta\,
d\phi^{2}).
\end{equation}
It is not clear what kind of material would be able
to support such a wormhole, given the zero density.
For that reason we will start the next section with
a black hole, whose main function is to generate the
gravitational field:
\begin{equation}\label{E:line3}
ds^{2}=-\left(1-\frac{2M}{r}\right)dt^{2}
+\frac{dr^2}{1-2M/r}+r^{2}(d\theta^{2}
+\text{sin}^{2}\theta\,d\phi^{2}).
\end{equation}
This approach turns out to be somewhat similar to the
modified black hole discussed in Refs.
\cite{mV96, DS07}:
\begin{equation}\label{E:line4}
ds^{2}=-\left(1-\frac{2M}{r}+\lambda^2\right)dt^{2}
+\frac{dr^2}{1-2M/r}+r^{2}(d\theta^{2}
+\text{sin}^{2}\theta\,d\phi^{2}).
\end{equation}
Here the main point is that for sufficiently small
$\lambda^2$, such a wormhole may be observationally
indistinguishable from a black hole.

%END OF SECTION

\section{Solutions}\label{S:solutions}
In this section we continue the theme of theoretically
constructing a wormhole based on a preexisting black
hole:
\begin{equation}\label{E:line5}
ds^{2}=-\left(1-\frac{2M}{r}\right)dt^{2}
+\frac{dr^2}{1-(2M+A)/r}+r^{2}(d\theta^{2}
+\text{sin}^{2}\theta\,d\phi^{2}), \quad A>0.
\end{equation}
(Observe that $r_0=2M+A$.)  In spite of our reliance
on a black hole, Eq. (\ref{E:line5}) does not represent
a thin shell in the sense of Visser \cite{mV96, PV95}.
Since the shape function is $b(r)=2M+A$, we have
$\rho(r)\equiv 0$ by Eq. (\ref{E:Einstein1}).  Since
$e^{2\Phi}=1-2M/r$, $\Phi=\frac{1}{2}\text{ln}\,
(1-2M/r)$ and we obtain from Eq. (\ref{E:Einstein2}),
\begin{equation}\label{E:pr}
   8\pi p_r=-\frac{2M+A}{r^3}+\frac{2M}{r^3}
   \frac{1-\frac{2M+A}{r}}{1-\frac{2M}{r}}.
\end{equation}
As a result, $\rho+p_r<0$ near the throat, so that
the null energy condition is indeed violated.  If
$A\rightarrow 0$, Eq. (\ref{E:pr}) shows that $p_r$
approaches $-1/8\pi r_0^2$, as one would expect.

The simple shape function also produces a simple
profile curve for the embedding surface \cite{MT88}.
From
\[
   \frac{dz}{dr}=\pm \left(\frac{r}{b(r)}
   -1\right)^{-1/2},
\]
we get the parabolas
\begin{equation}
    z(r)=\pm 2(2M+A)\sqrt{\frac{r}{2M+A}-1}.
\end{equation}

For the tidal acceleration at the throat \cite{MT88},
\begin{equation}\label{E:tidal1}
    a_T=\left|\frac{b'r-b}{2r^2}\Phi'\Delta\xi\right|
\end{equation}
in terms of some distance $\Delta\xi$.  (In Ref.
\cite{MT88}, $\Delta\xi=2\,\text{m}$, the approximate
height of a person.)  Eq. (\ref{E:tidal1}) yields
\begin{equation}\label{E:tidal2}
   a_T=\left|\frac{M}{2(2M+A)^3\left(1-
   \frac{2M}{2M+A}\right)}\Delta\xi\right|.
\end{equation}

\subsection{Small $A$}\label{S:smallA}
In line element (\ref{E:line5}), if $A$ is a small
constant, we are dealing with a ``black-hole mimicker,"
just as in line element (\ref{E:line4}).  According to
Eq. (\ref{E:tidal2}), the size of the radial tidal force
depends on the size of the region of high tension.  For
the right choice of $A$, $a_T$ can be made to match the
tidal acceleration of the black hole itself.  (This
value would be the smallest value allowed in our
wormhole construction.)

\subsection{Large $A$}
A sufficiently large $A$ will result in a wormhole with
low tidal forces.  For example, suppose the black hole
has one solar mass, i.e., $M=1474\,\text{m}$.  Its event
horizon is the sphere $r=2948\,\text{m}$.  If
$A=5000\,\text{km}$, then
\[
    a_T<(10^8\,\text{m})^{-2},
\]
the criterion for human traversability suggested in Ref.
\cite{MT88}.  Similar results can be obtained for any
preexisting black hole.
%END OF SECTION

\section{$A=A(t)$}
In this section we are going to seek another
connection to the preexisting black hole by
replacing $A$ by the time-dependent function
$A=A(t)$.

Possible connections between black holes and wormholes
have been the subject of investigations for some time.
Thus Hayward \cite{sH02} proposed a unified framework
for black holes and traversable wormholes, a synthesis
that makes them essentially interconvertible.
Kardashev et al. \cite{KNS07} considered the
possibility that compact astrophysical objects such
as active galactic nuclei may be current or former
entrances to wormholes.  Kuhfittig \cite{pK08}
suggested that if the Universe had indeed crossed
the phantom divide, as noted by some researchers
\cite{SS06}, then wormholes could have formed
naturally and some of these wormholes could have
become black holes at a time closer to the present.
Some could have become quasi-black holes, as
defined by Lemos and Zaslavskii \cite{LZ07}.

The conversion of a black hole to a wormhole is
made explicit by Hayward \cite{sH02a}.  A toy model
that generalizes the CGHS two-dimensional dilaton
gravity model by including a ghost scalar field made
it possible to describe explicitly the evolution of
the black hole to a wormhole: the ghost radiation,
which acts like exotic matter, causes the event
horizons of the initial black hole to become
timelike and may eventually merge to form the throat
of the wormhole.  In that event, the trapped region
would have evaporated, so that the singularity
itself would have disappeared.

Sec. \ref{S:solutions} suggests another toy model
which, with Ref. \cite{sH02a} taken into account,
results in a different kind of wormhole, more akin
to the thin-shell wormhole in Ref. \cite{PV95}.
Replacing the constant $A$ in Eq. (\ref{E:line5})
by the time-dependent function $A=A(t)$, we have
\begin{multline}\label{E:line6}
ds^{2}=-\left(1-\frac{2M}{r}\right)dt^{2}
+\frac{dr^2}{1-[2M+A(t)]/r}+r^{2}(d\theta^{2}
+\text{sin}^{2}\theta\,d\phi^{2}),\\
\quad A(t)>0\quad \text{for}\quad t>0.
\end{multline}
Since the starting point is a black hole, we need
to assume that $A(0)=0$.  This model has the
advantage of being easy to analyze while still
managing to produce results that are both
interesting and physically plausible.  Whenever
$A(t)$ is close to zero, we are momentarily back
to Subsection \ref{S:smallA}.  So $A(t)$ has to
assume larger values, as well, in order to produce
a regular wormhole instead of just a black-hole
mimicker.

According to Ref. \cite{sH02a}, the most important
physical consequence is that the two event horizons
have now become timelike, thereby forming two throats,
one on each side of the center.  It is shown in Ref.
\cite{HV} that this is to be expected since our
wormhole is dynamic.  Moreover, returning to the
initial black hole, let us recall that, according
to Ref. \cite{MTW}, page 838, a spacelike hypersurface
extending from a region in one universe to that of
another is not static.  The reason is that the
``bridge" enlarges and contracts rapidly, so rapidly,
in fact, that not even a light ray can pass through.
Line element (\ref{E:line6}) could model this behavior
for a proper choice of $A(t)$ (with the understanding
that the event horizons are now throats), as well as
the continuing evolution of the structure as ever more
exotic matter is added.  This structure is best
described by the Penrose diagram in Fig. 1.  Even
\begin{figure}[tbp]
\begin{center}
\includegraphics[width=0.8\textwidth]{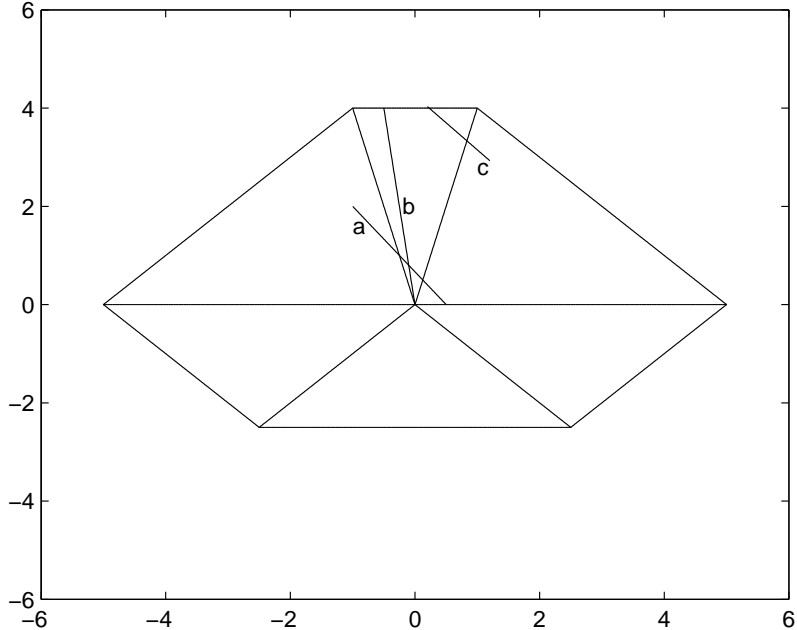}
\end{center}
\caption{The Penrose diagram.}
\end{figure}
though it is now easy to cross the two throats following
a timelike geodesic, (line $a$), the singularity is still
present.  Judging from lines $b$ and $c$, however, the
only way that a traveler can crash into the singularity
is by either traveling too slowly or by delaying the
start of the trip for too long in any given cycle.  In
other words, the rates of expansion and contraction have
slowed enough to enable the traveler to avoid the
singularity, or, what amounts to the same thing, $A(t)$
in Eq. (\ref{E:line6}) has a longer period.  Continuing
this process, the two throats could eventually merge to
form the throat of a static wormhole, while the
singularity is no longer part of the wormhole spacetime.
So Fig. 1 represents a transitional structure between a
black hole and a wormhole.

It is conceivable that Fig. 1 represents an end
result, also mentioned in Ref. \cite{sH98}, in which
case we are dealing with a new kind of wormhole, one
that could be described as a black hole that
periodically contracts and pinches off before
expanding again, but doing so slowly enough to
permit passage.

If the expansions and contractions stop entirely,
then the result is a static wormhole.  From the shape
function $b(r)=2M+A(t)$, with $A(t)$ fixed, we still
have $\rho(r)\equiv 0$, as in Sec. \ref{E:introduction}.
But some of the exotic matter would have ended up on
the throat, so that the density is best described by
the delta-function form $\rho (r)
=A\,\delta (r-r_0)$.  The throat is therefore a thin
shell that contains all of the exotic matter.  The
singularity has been excised from the wormhole
spacetime, thereby avoiding a naked singularity.
The resulting wormhole differs somewhat from the
thin-shell wormhole in Ref. \cite{PV95}, which is
assumed to have a throat outside the event horizon
of the black hole.  In our wormhole, the event
horizon itself has been converted to a throat.

%END OF SECTION

\section{$A=A(r)$}

In this section we consider the case where $A$ in line
element (\ref{E:line5}) is a function of $r$ alone,
i.e., $A=A(r)$, representing a static wormhole.  In
other words, the conversion from a black hole to a
wormhole has already been completed.  To obtain
$A(r)$, we are going to assume a specific model for
the exotic matter.

Given that the throat of the wormhole would be a
considerable distance away from the original
singularity, $p_r$ is relatively small, according
to Eq. (\ref{E:pr}).  So to model the exotic matter,
we take the equation of state to be $p_r=\omega\rho$,
$\omega <-1$.  In other words, the exotic matter is
similar to phantom dark energy.  Since
$b(r)=2M+A(r)$,
\begin{equation}
   8\pi p_r=8\pi\omega\rho=\omega\frac{A'(r)}{r^2}=
   -\frac{2M+A(r)}{r^3}+\frac{2M}{r^3}
   \frac{1-\frac{2M+A(r)}{r}}{1-\frac{2M}{r}}
\end{equation}
and
\begin{equation}
   \omega A'(r)=
   -\frac{2M+A(r)}{r}+\frac{2M}{r}
   \frac{1-\frac{2M+A(r)}{r}}{1-\frac{2M}{r}}.
\end{equation}
This equation has the following simple solution:
\begin{equation}
   A(r)=c(r-2M)^{-\frac{1}{\omega}}, \quad \omega<-1,
\end{equation}
where $c>0$ is an arbitrary constant.  It is
interesting to note that $A(2M)=0$ for all $c$.
Furthermore,
\[
   \frac{b(r)}{r}\rightarrow 0\,\, \text{as}\,\,
     r\rightarrow \infty
\]
if, and only if, $\omega<-1$, so that the
wormhole spacetime is asymptotically flat.

The location of the throat $r=r_0$ depends on the
constant $c$ (in addition to $\omega$): since
$b(r_0)=r_0$, we can find $r_0$ by solving the
equation $b(r)-r=0$.  So from
\[
    2M+c(r-2M)^{-\frac{1}{\omega}}-r=0,
\]
we obtain
\begin{equation}\label{E:newthroat}
    r_0=2M+c^{\frac{1}{1/\omega+1}}.
\end{equation}
It now follows that
\begin{equation}
    b'(r_0)=-\frac{1}{\omega}<1,\quad
    \text{since}\quad \omega<-1.
\end{equation}
So we are getting a well-behaved wormhole for
every $c>0$.

Returning to the shape function, from
Eq. (\ref{E:Einstein1}),
\begin{equation*}
     b(r)=b(r_0)
       +\int^r_{r_0}8\pi\rho(r')(r')^2dr'=2m(r).
\end{equation*}
Then the mass inside a sphere of radius $r$ is given by
\begin{equation}\label{E:mass}
      \frac{1}{2}b(r)=M+\frac{1}{2}c(r-2M)
      ^{-\frac{1}{\omega}},
\end{equation}
which is the mass $M$ of the black hole plus the
additional matter needed to convert the black hole to
a wormhole.  Since the conversion is completed,
Eq. (\ref{E:mass}) suggests that the singularity has
been dissolved to become part of the wormhole's mass.

The equation of state $p_r=\omega\rho$, $\omega<-1$,
suggests that we are, in fact, dealing with phantom
dark energy.  So if the black hole continues to draw
in phantom energy from the cosmological background,
the result may be a wormhole, a possibility already
considered by Hayward \cite{sH02a} for the simple
reason that this could provide a solution to the
information paradox.
%END OF SECTION

\section{Conclusion}
The first part of this paper discusses the theoretical
construction of a particularly simple class of traversable
wormholes having zero density.  Since the density of
matter is not ordinarily zero, we rely for our
construction on an already existing black hole.  One
of the parameters in the line element can in principle
be adjusted to yield a whole set of solutions ranging
from black-hole mimickers to low-tidal-force wormholes.
A toy model suggested by the first model then discusses
the possible conversion of the black hole to a wormhole
by adding exotic matter, as discussed in Ref.
\cite{sH02a}.  According to our toy model, the result
is either a time-dependent wormhole with two throats
that periodically expands and contracts slowly enough 
to permit passage or a plausible transitional structure 
between a black hole and the resulting wormhole.  The 
singularity is still present in both cases.  If the 
end result is a static wormhole, it must be viewed 
as a thin-shell wormhole with the original singularity 
excised from the wormhole spacetime, thereby avoiding 
a naked singularity.  Assuming a specific equation of
state for the exotic matter yields a solution suggesting
that the singularity of the black hole dissolves to
become part of the mass of the wormhole.


\begin{thebibliography}{20}

\bibitem{MT88}M.S. Morris and K.S. Thorne, ``Wormholes in
   spacetime and their use for interstellar travel: A tool
   for teaching general relativity," Amer. J. Phys.
   \textbf{56}, 395 (1988).
\bibitem{sH02}S.A. Hayward, ``Black holes and traversible
   wormholes: a synthesis," arXiv: 0203051.
\bibitem{sH98}S.A. Hayward, ``Dynamic wormholes," arXiv: gr-qc/9805019.
\bibitem{mV96}M. Visser, Lorentzian wormholes: from
   Einstein to Hawking, New York, Springer-Verlag, 1996.
\bibitem{DS07}T. Damour and S.N. Solodukhin, ``Wormholes
   as black hole foils," Phys. Rev. D \textbf{76},
   024016 (2007).
\bibitem{PV95}E. Poisson and M. Visser, ``Thin-shell
   wormholes: Linearized stability," Phys. Rev. D
   \textbf{52}, 7318 (1995).
\bibitem{KNS07} N.S. Kardashev, I.D. Novikov, and A.A.
   Shatskiy, ``Astrophysics of wormholes,"  Int. J. Mod.
   Phys. \textbf{16}, 909 (2007).
\bibitem{pK08}P.K.F. Kuhfittig, ``Could some black holes
   have evolved from wormhoes?" Schol. Res. Exch.
   \textbf{2008}, 296158 (2008).
\bibitem{SS06}J. Sola and H. Stefancic, ``Dynamical dark
   energy or variable cosmological parameters?" Mod. Phys.
   Lett. A \textbf{21}, 479 (2006).
\bibitem{LZ07}J.P.S. Lemos and O.B. Zaslavskii,
   ``Quasi-black holes: definition and general properties,"
   Phys. Rev. D \textbf{76}, 084030 (2007).
\bibitem{sH02a}S.A. Hayward, ``Dilaton wormholes:
   construction, operation, maintainance and collapse
   to black holes," Phys. Rev. D \textbf{65}, 064003
   (2002).
\bibitem{HV}D. Hochberg and M. Visser, ``Dynamic wormholes,
   anti-trapped surfaces, and energy conditions," Phys.
   Rev. D \textbf{58}, 044021 (1998).
\bibitem{MTW}C.W. Misner, K.S. Thorne, and J.A. Wheeler,
   Gravitation, San Francisco, Freeman, 1973.

 \end{thebibliography}
\end{document}